\documentclass[reprint,twocolumn,superscriptaddress,pre,amsmath,amssymb,aps]{revtex4-1}

\usepackage{natbib}	
\usepackage{graphicx,color,rotating}
\usepackage[latin1]{inputenc}
\usepackage{textcomp}
\usepackage{dcolumn}
\usepackage{bm}     
\usepackage{upgreek}
\usepackage{subdepth}  			
\usepackage{multirow}
\usepackage[normalem]{ulem}     
\usepackage{float}

\usepackage{array}
\newcolumntype{L}[1]{>{\raggedright\let\newline\\\arraybackslash\hspace{0pt}}m{#1}}
\newcolumntype{C}[1]{>{\centering\let\newline\\\arraybackslash\hspace{0pt}}m{#1}}
\newcolumntype{R}[1]{>{\raggedleft\let\newline\\\arraybackslash\hspace{0pt}}m{#1}}
\usepackage{ulem} 

\usepackage{hyperref}						
\usepackage{xcolor}						
\hypersetup{							
    colorlinks,
    linkcolor={blue!80!black},
    citecolor={blue!80!black},
    urlcolor={blue!80!black}
}
\newcommand{\SImum}{\textrm{\textmu{}m}}
\newcommand{\SImuL}{\textrm{\textmu{}L}}
\newcommand{\SICel}{^\circ\!\textrm{C}}
\newcommand{\eqlab}[1]{\label{eq:#1}}
\renewcommand{\eqref}[1]{Eq.~(\ref{eq:#1})}
\newcommand{\eqnoref}[1]{(\ref{eq:#1})}

\newcommand{\secref}[1]{Section~\ref{sec:#1}}

\newcommand{\seclab}[1]{\label{sec:#1}}

\begin{document}

\title{Particle-size-dependent acoustophoretic motion and depletion of micro- and nanoparticles at long time scales}

\author{Wei Qiu}
\email{wei.qiu@bme.lth.se}
\affiliation{Department of Physics, Technical University of Denmark, DTU Physics Building 309, DK-2800 Kongens Lyngby, Denmark}
\affiliation{Department of Biomedical Engineering, Lund University, Ole R\"{o}mers v\"{a}g 3, 22363, Lund, Sweden}

\author{Henrik Bruus}
\email{bruus@fysik.dtu.dk}
\affiliation{Department of Physics, Technical University of Denmark, DTU Physics Building 309, DK-2800 Kongens Lyngby, Denmark}

\author{Per Augustsson}
\email{per.augustsson@bme.lth.se}
\affiliation{Department of Biomedical Engineering, Lund University, Ole R\"{o}mers v\"{a}g 3, 22363, Lund, Sweden}

\date{21 June 2020}

\begin{abstract}
We present three-dimensional measurements of size-dependent acoustophoretic motion of microparticles with diameters from 4.8~$\SImum$ down to 0.5~$\SImum$ suspended in either homogeneous or inhomogeneous fluids inside a glass-silicon microchannel and exposed to a standing ultrasound wave. To study the cross-over from radiation force dominated to streaming dominated motion as the particle size is decreased, we extend previous studies to long time scales, where the particles smaller than the cross-over size move over distances comparable to the channel width.  We observe a particle-size-dependent particle depletion at late times for the particles smaller than the cross-over size. The mechanisms behind this depletion in homogeneous fluids are rationalized by numerical simulations which take the Brownian motion into account. Experimentally, the particle trajectories in inhomogeneous fluids show focusing in the bulk of the microchannel at early times, even for the particles below the critical size, which clearly demonstrates the potential to manipulate submicrometer particles.
\end{abstract}

\maketitle


\section{Introduction}

Microscale acoustofluidics, relying on ultrasound-induced forces, has emerged as a tool in contemporary biotechnology to concentrate~\cite{Nordin2012, Carugo2014}, trap~\cite{Evander2012, Collins2015}, wash~\cite{augustsson2012a}, align~\cite{Manneberg2008a}, synchronize~\cite{Thevoz2010}, and separate suspended cells~\cite{Petersson2007, Augustsson2012, Ding2014} in a flow-through format. The acoustophoretic motion of suspended microparticles in a microchannel exposed to a standing acoustic wave is dominated by two forces: The acoustic radiation force from the scattering of sound waves on the particles and the Stokes drag force from the vortical acoustic streaming induced in the fluid by the acoustic wave. The acoustic radiation force is directed along the gradient of the acoustic potential and is often used to move particles in the lateral cross section of the channel in a controlled manner. In contrast, the acoustic streaming typically leads to mixing of particles inside the microchannel, and it has only been applied for particle manipulation in some special cases~\cite{Wiklund2012, Hammarstrom2012, Antfolk2014, Mao2017}.

It is well established experimentally and theoretically that the particle velocity induced by the acoustic radiation force in a homogeneous fluid scales with the square of the particle radius~\cite{Barnkob2012a, Muller2012, Muller2013}. Therefore, there exists a cross-over particle size above which the particle velocity is dominated by the radiation force and below which it is dominated by the acoustic streaming. For polystyrene particles suspended in water and exposed to a 2-MHz standing half-wave, this cross-over size is observed to be around 2~$\SImum$~\cite{Barnkob2012a, Muller2013}, as predicted by a simple scaling law equating the velocity from the two contributions~\cite{Muller2012}. However, the numerical work underlying this analysis is limited to particle trajectories and particle distributions on the short time scale less than 10~s after the onset of sound with a moderate acoustic energy density~\cite{Muller2012}. In that case, most of the particles below the cross-over size only move short distances compared to the length scale of the channel cross section, and it is therefore very likely that the full picture of particle trajectories is not revealed. Previous experimental work measured two-dimensional (2D) particle motion for many particle sizes~\cite{Barnkob2012a}, in a horizontal $x$-$y$ plane at channel mid-height. In contrast, three-dimensional (3D) measurements, which allows mapping of the vertical plane in which boundary-driven Rayleigh streaming typically occurs, has only been conducted for particle diameters far below (0.5~$\SImum$) or above (5~$\SImum$) the cross-over diameter~\cite{Muller2013}. Consequently, the acoustophoretic motion in homogeneous fluids of particles with sizes close to the cross-over diameter remains unclear. Considering that some of the target bio-particles (such as blood platelets and bacteria) to be manipulated using acoustofluidic devices are in a size-range close to the cross-over size, it is essential to understand thoroughly the physics of the 3D acoustophoretic particle motion for various particle sizes on the long time scale.

Similarly, the acoustophoretic motion of particles suspended in inhomogeneous fluids has not been fully studied since their recent introduction to acoustofluidics. It was discovered that in acoustofluidic systems filled with fluids that are made spatially inhomogeneous in density and compressibility, an acoustic body force arises which leads to fluid relocation or stabilization, and to the suppression of acoustic streaming~\cite{Deshmukh2014, Augustsson2016, Karlsen2016, Karlsen2018, qiu2019experimental}. Based on this phenomenon, iso-acoustic focusing was developed to achieve size-independent particle separation and measure the acoustic properties of cells according to their equilibrium positions after migration in a fluid with a smooth acoustic impedance profile~\cite{Augustsson2016}.  Later, we showed how the suppressed streaming gradually reappears as the inhomogeneity slowly disappears due to diffusion and advection~\cite{Karlsen2018, qiu2019experimental}. The streaming suppression results in a significant change of the cross-over diameter of suspended particles from streaming dominated to radiation force dominated motion, indicating a clear potential to manipulate particles in the sub-micrometer size range.

In this paper, we extend previous experimental studies of particle-size-dependent acoustophoretic motion of particles suspended in homogeneous and inhomogeneous fluids to long time scales, for which the particles below the cross-over size move distances comparable to the size of the channel cross section, and we characterize the associated cross-over from radiation force dominated to streaming dominated behaviour. The results are obtained by the use of particle tracking velocimetry on suspended particles in an ultrasound half-wavelength silicon-glass microchip with a rectangular cross section. In \secref{theory}, we give a brief summary of the theory of acoustophoretic motion in homogeneous and inhomogeneous fluids, followed in \secref{experiments} by a description of the experimental setup, materials, and methods used to measure particle trajectories in homogeneous and inhomogeneous mediums. In \secref{exp_homog}, we present the experimental results for acoustophoretic motion at long time scales in homogeneous fluids, identifying particle-size-dependent depletion regions as part of the cross-over behavior. The mechanisms behind the particle-size-dependent particle depletion, including the effects of Brownian motion, are analyzed numerically and discussed in \secref{discus_homog}. Finally, in \secref{exp_inhomog},  we show experimentally, that also for  inhomogeneous fluids, the acoustophoretic motion at long time scales leads to particle depletion in particle-size-dependent regions. We present our conclusions regarding the cross-over from radiation force dominated to streaming dominated behavior at long time scales in \secref{conclusion}.

\section{Theoretical background}
\seclab{theory}

\subsection{Time-averaged acoustic forces on a single spherical microparticle}

A suspended single spherical microparticle in an acoustic field experiences an acoustic radiation force $\bm{F}^\mathrm{rad}$ due to the scattering of acoustic waves on the particle and the Stokes drag force $\bm{F}^\mathrm{drag}$ due to the acoustic streaming. The time-averaged acoustic radiation force $\bm{F}^\mathrm{rad}$ on a freely suspended single spherical particle of radius $a$, density $\rho_\mathrm{p}$, and compressibility $\kappa_\mathrm{p}$ in a viscous fluid is given by \cite{Settnes2012}
\begin{equation}
\eqlab{Frad}
\bm{F}^\mathrm{rad} = -\pi a^3
\bigg\{\frac{2}{3}\kappa_0\mathrm{Re}\big[{f_0}{p_1}\boldsymbol{\nabla} p^{*}_1\big]
- \rho_0\mathrm{Re}\big[{f_1}{\boldsymbol{v}_1}\cdot \boldsymbol{\nabla} \boldsymbol{v}^{*}_1\big]\bigg\},
\end{equation}
where the asterisk denotes complex conjugation. The monopole and dipole scattering coefficients $f_0$ and $f_1$ are given by
 \begin{align}
 f_0(\widetilde \kappa ) = 1 - \widetilde \kappa, \quad
 f_1(\widetilde \rho, \widetilde \delta)
 &= \frac{2\big[1-\Gamma \big(\widetilde \delta \big)\big]\big(\widetilde \rho -1\big)}{2 \widetilde \rho+1-3 \Gamma \big(\widetilde \delta \big)},
 \end{align}
where $\Gamma \big(\widetilde \delta \big) = -\frac{3}{2} \big[1+\mathrm{i} \big(1+\widetilde \delta \big)\big] \widetilde \delta$, with $\widetilde \kappa = \frac{\kappa_\mathrm{p}}{ \kappa_0}$, $\widetilde \rho = \frac{\rho_\mathrm{p}}{\rho_0}$, and $\widetilde \delta = \frac{\delta}{a}$. For the case of a half-wavelength standing-wave field in $y$ direction, the expression for $\bm{F}^\mathrm{rad}$ simplifies to
 \begin{subequations}
 \label{1Dradiationforce}
 \begin{align}
 \bm{F}^\mathrm{rad}\big(y \big) &= 4 \pi a^\mathrm{3} k_\mathrm{y} E_\mathrm{ac} \Phi \big(\widetilde \kappa, \widetilde \rho, \widetilde \delta \big) \sin(2k_y y),
 \\
 \Phi \big(\widetilde \kappa, \widetilde \rho, \widetilde \delta \big)
 &= \frac{1}{3}f_0\big(\widetilde \kappa \big)+\frac{1}{2}\mathrm{Re}\big[f_1\big(\widetilde \rho, \widetilde \delta \big)\big],
\end{align}
\end{subequations}
where $\Phi$ is the acoustic contrast factor.

The time-averaged Stokes drag force $\bm{F}^\mathrm{drag}$ on a spherical particle moving with velocity $\bm{u}$ in a fluid having the streaming velocity $\big\langle\bm{v}_2\big\rangle$ is given by the usual expression
\begin{equation}
\label{Dragforce}
\bm{F}^\mathrm{drag} = 6\pi\eta_0 a \big( \big\langle\bm{v}_2\big\rangle - \bm{u} \big),
\end{equation}
which is valid for particles sufficiently far from the channel walls \cite{Koklu2010}.

\subsection{Boundary-driven acoustic streaming in a fluid}

The theory of steady acoustic streaming in a homogeneous fluid inside an infinite parallel-plate channel was given analytically by Lord Rayleigh~\cite{LordRayleigh1884} in the isothermal case for a channel of height $H$ with its two plates placed symmetrically around the $x$-$y$ plane at $z = \pm\frac12 H$ and with an imposed standing acoustic wave of wavelength $\lambda$ and wavenumber $k_y=2\pi/\lambda$ along the $y$ direction $\bm{e}_y$: $p_1(y) = p_\mathrm{ac} \sin(k_y y)\:\mathrm{e}^{-\mathrm{i}\omega t}$ and $\bm{v}_1 = v_\mathrm{ac}\cos(k_y y) \:\mathrm{e}^{-\mathrm{i}\omega t}\:\bm{e}_y$. In the case of  $\lambda \gg H \gg \delta$, where $\delta = \sqrt{2\eta_0/(\rho_0\omega)}$ is the thickness of the viscous boundary layer, Rayleigh found the time-averaged components $\big\langle v_{2y} \big\rangle$ and $\big\langle v_{2z}\big\rangle$ of the streaming velocity $\big\langle\bm{v}_2(y,z) \big\rangle$ outside the viscous boundary layer to be
\begin{subequations}
\label{2Dstreamingfield}
\begin{align}
\big\langle v_{2y} \big\rangle&= \frac{3}{8}\frac{v^2_\mathrm{ac}}{c_0}\sin\left(2k_y y\right) \left[1-3\frac{(2z)^2}{H^2}\right]\frac{1}{2},\\
\big\langle v_{2z} \big\rangle&= \frac{3}{8}\frac{v^2_\mathrm{ac}}{c_0}\cos\left(2k_y y\right) \left[\frac{(2z)^3}{H^3} - \frac{2z}{H} \right]\frac{kH}{2}.
\end{align}
\end{subequations}
The amplitudes $p_\mathrm{ac}$ and $v_\mathrm{ac}$ are both related to the acoustic energy density $E_\mathrm{ac}$,
 \begin{equation}
 E_\mathrm{ac} = \frac14 \rho_0 v_\mathrm{ac}^2 = \frac14 \kappa_0 p_\mathrm{ac}^2,
 \end{equation}
where $\kappa_{0}$ is the medium compressibility. This solution of the ideal case captures the main feature in the streaming flow, in particular the characteristic quadrupolar flow roll structure. For a closed rectangular channel with no-slip walls, an analytical solution for homogeneous fluids exists~\cite{Muller2013}, but none is known for inhomogeneous fluids. In our numerical simulations below, we compute the streaming velocity numerically in all cases.

\subsection{The acoustic body force acting on inhomogeneous fluids}

As we have shown in Ref.~\cite{Karlsen2016}, the acoustic fields acting on the short time scale $t$ give rise to an acoustic body force $\bm{f}_\mathrm{ac}$ acting on the inhomogeneous fluids on the slow time scale $\tau$. This body force is derived from the nonzero divergence in the time-averaged (over one oscillation period $2\pi/\omega$) acoustic momentum-flux-density tensor $\big\langle\bm{\Pi}\big\rangle$,
\begin{equation}
\label{Bodyforce}
\bm{f}_\mathrm{ac} = -\boldsymbol{\nabla}\cdot\big\langle\bm{\Pi}\big\rangle.
\end{equation}
The second-order quantity $\big\langle\bm{\Pi}\big\rangle$ is given by products of the first-order acoustic fields $p_1$ and $\bm{v}_1$~\cite{Landau1993},
\begin{equation}
\big\langle\bm{\Pi}\big\rangle = \big\langle{p_2}\big\rangle \mathbf{I} + \big\langle{{\rho_0}{\bm{v}_1}{\bm{v}_1}}\big\rangle,
\end{equation}
where the time-averaged, second-order mean Eulerian excess pressure $\big\langle{p_2}\big\rangle$ takes the form
\begin{equation}
\big\langle{p_2}\big\rangle = \frac{1}{4} \kappa_0 |p_1|^2 - \frac{1}{4} \rho_0 |\bm{v}_1|^2.
\end{equation}
The acoustic body force $\bm{f}_\mathrm{ac}$ in Eq.~(\ref{Bodyforce}) was derived on the slow hydrodynamic time scale $\tau$ in Ref.~\cite{Karlsen2016} in terms of continuous spatial variations in the fluid density $\rho_0$ and compressibility $\kappa_0$, or equivalently in density $\rho_0$ and sound speed $c_0$,
 \begin{subequations}
 \eqlab{facFull}
 \begin{align}
 \bm{f}_\mathrm{ac} &= - \frac{1}{4} |p_1|^2 \boldsymbol{\nabla}\kappa_0
 - \frac{1}{4} |\bm{v}_1|^2 \bm{\nabla}\rho_0
 \\
 &= \frac14 \big(\kappa_0 |p_1|^2 - \rho_0|\bm{v}_1|^2\big)\frac{\bm{\nabla}\rho_0}{\rho_0}
 + \frac12 \kappa_0 |p_1|^2 \frac{\bm{\nabla}c_0}{c_0}.
 \end{align}
 \end{subequations}
The acoustic body force relocates regions in the fluid of different density and compressibility~\cite{Karlsen2016} and it suppresses boundary-driven streaming~\cite{Karlsen2018}. Regarding the acoustic radiation force on a particle suspended in an inhomogeneous fluid, the expression~\eqnoref{Frad} for $\bm{F}^\mathrm{rad}$ in terms of the local acoustic pressure $p_1$ and velocity $\bm{v}_1$ remains valid, as long as the particle is much smaller than both the acoustic wavelength and the distance over which the properties of the fluid change. The latter condition ensures that waves scattered off a particle does not back-scatter to the particle due to inhomogeneities.

\section{Experiments}
\seclab{experiments}

\subsection{Experimental setup and materials}
\begin{figure}[!b]
\centering
\includegraphics[width=1.0\columnwidth]{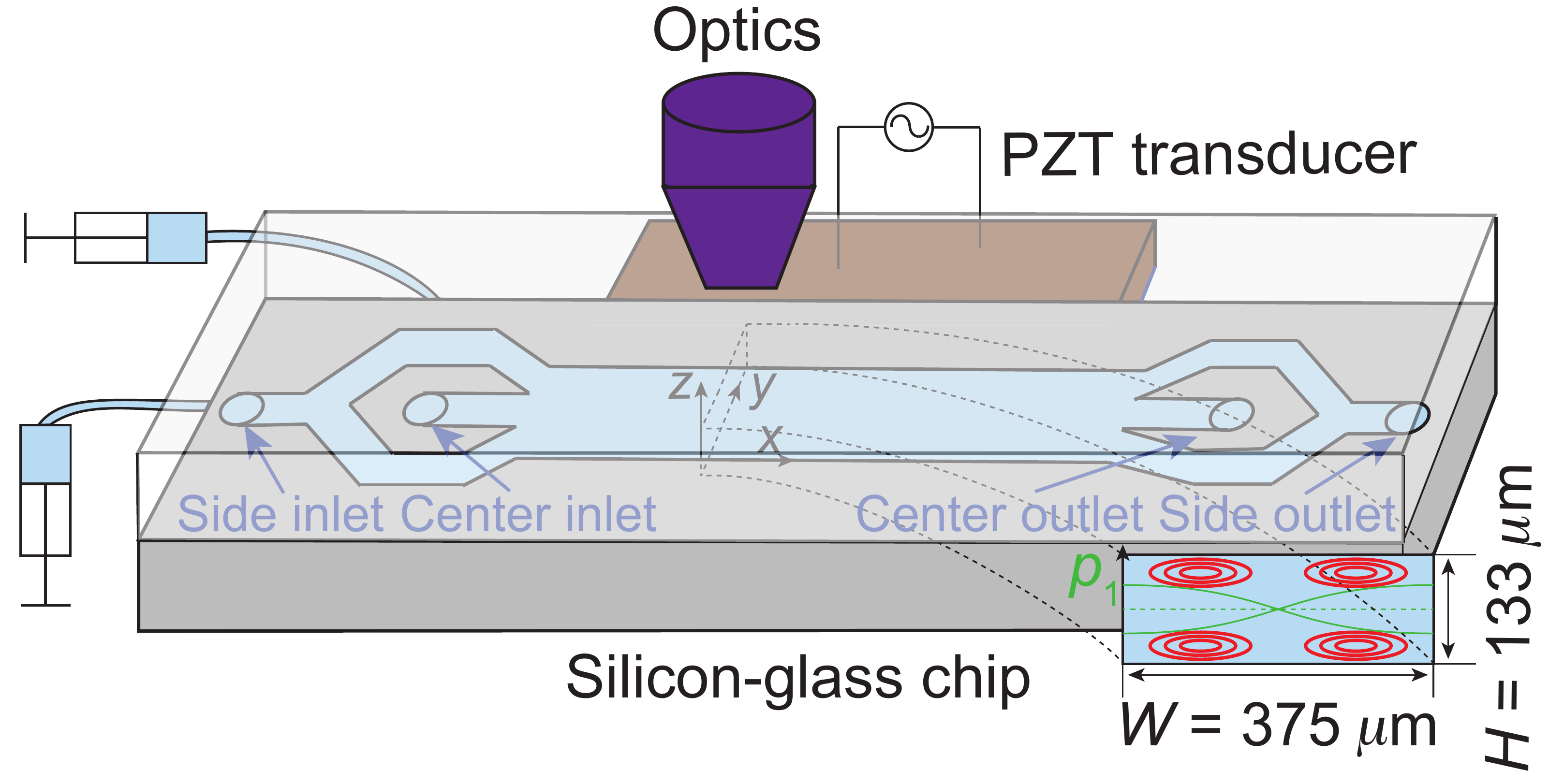}
\caption[]{\label{Sketch}
Sketch of the acoustofluidic silicon chip (light gray) sealed with a glass lid, which allows optical recording (dark purple) of the tracer bead motion (inset, dark red trajectories) in the channel cross section. The piezoelectric transducer (dark brown) excites the resonant half-wave pressure field $p_1$ (inset, light green) at 2 MHz.}
\end{figure}
The microchip and setup used for this experiment, the same as in Ref.~\cite{qiu2019experimental}, are briefly summerized here and in Fig.~\ref{Sketch}. The silicon chip consists of a straight channel of length $L = 24$~mm, width $W = 375~\SImum$, and height $H = 133~\SImum$, resulting in the aspect ratio $\alpha = 0.355$. The chip was sealed by a 1-mm-thick Pyrex lid by anodic bonding, and an 18~mm $\times$ 6.4~mm $\times$ 1.0~mm lead zirconate titanate (PZT) transducer (Pz26, $\mathrm{Ferroperm^{TM}}$ Piezoceramics, Meggitt A/S, Denmark) was bonded underneath using cyanoacrylate glue (Loctite Super Glue, Henkel Norden AB, Stockholm, Sweden). At the center inlet, three flow streams join in a trifurcation of which the two side streams are routed via a common port whereas the center stream has a separate port. At the end of the channel, the outlet has the same trifurcated configuration as the inlet. Pieces of silicone tubing (outer diameter 3~mm, inner diameter 1~mm, and length 7~mm) were glued to the chip inlets and outlets. The inlet flow streams are routed via a motorized four-port, two-way diagonal valve so that the flow can be stopped by a short-circuit of the side inlet with the center inlet, and the outlet stream is routed via a two-port solenoid valve for stopping the flow. The inlets were used for injecting the same fluid everywhere to obtain a homogeneous medium, or for injecting two different liquids to obtain an inhomogeneous medium, whereas only one outlet was used for collecting the waste, while the other outlet was blocked during all measurements. A Pt100 thermo-resistive element was bonded to the PZT transducer to record the temperature. The PZT transducer was driven by a function generator (AFG3022B, Tektronix, Inc., Beaverton, Oregon, USA), and the waveforms of the applied voltages to the transducer were monitored by an oscilloscope (TDS1002, Tektronix, Inc., Beaverton, Oregon, USA).  The liquids were injected in the channel using syringe pumps (neMESYS, Cetoni GmbH, Korbussen, Germany) controlled by a computer interface.

Fluorescent green polystyrene beads with nominal diameters $d$ of 0.5 $\SImum$, 1.0 $\SImum$, 1.9 $\SImum$, 3.1 $\SImum$, and 4.8 $\SImum$ (Molecular Probes, Thermo Fisher Scientific, Waltham, MA, USA) were used to study the size-dependent particle trajectories in homogeneous and inhomogeneous fluids. In the homogeneous case, 2.5\% Ficoll PM70 (GE Healthcare Biosciences AB, Uppsala, Sweden) was used as the medium; whereas in the inhomogeneous case, 5\% Ficoll PM70 and Milli-Q water were laminated in the center and side streams, respectively. All relevant material properties are listed in Table~\ref{MaterialParameters}, and the detailed material properties of Ficoll PM70 can be found in Ref.~\cite{qiu2019experimental}.
\renewcommand\arraystretch{1.1}
\begin{table}[!t]
\caption{\label{MaterialParameters} Material parameters for polystyrene and 2.5\% Ficoll PM70 at temperature $T = 25~\SICel$. The mechanical properties of polystyrene were taken from the literatures, whereas the diffusion coefficient $D$ for each size of particles was calculated from the Stokes-Einstein equation. The parameters for 2.5\% Ficoll PM70 were measured experimentally.}
\begin{ruledtabular}
\begin{tabular}{ l r l l r r}
Polystyrene \\ \noalign{\smallskip}  \hline \noalign{\smallskip}
Density &  $\rho_\mathrm{ps}$ & 1050 & $\mathrm{kg\,m^{-3}}$ \\
Longitudinal speed of sound & $c_\mathrm{ps}$ & 2350 & $\mathrm{m\,s^{-1}}$ \\
Poisson's ratio & $\sigma_\mathrm{ps}$ & 0.35 & \\
Compressibility & $\kappa_\mathrm{ps}$ & 249 & $\mathrm{TPa^{-1}}$ \\
Diffusion coefficient & $D_\mathrm{0.5}$ & $7.46\,\times10^{-13}$ & $\mathrm{m^{2}\,s^{-1}}$ \\
				     	& $D_\mathrm{1.0}$ & $3.73\,\times10^{-13}$ & $\mathrm{m^{2}\,s^{-1}}$ \\
				     	& $D_\mathrm{1.9}$ & $1.96\,\times10^{-13}$ & $\mathrm{m^{2}\,s^{-1}}$ \\
					& $D_\mathrm{3.1}$ & $1.20\,\times10^{-13}$ & $\mathrm{m^{2}\,s^{-1}}$ \\
					& $D_\mathrm{4.8}$ & $0.78\,\times10^{-13}$ & $\mathrm{m^{2}\,s^{-1}}$ \\ \noalign{\smallskip} \hline \noalign{\smallskip}
2.5\% Ficoll PM70 \\ \noalign{\smallskip} \hline \noalign{\smallskip}
Density & $\rho_0$ & 1005.6 & $\mathrm{kg\,m^{-3}}$ \\
Longitudinal speed of sound & $c_0$ & 1502.5 & $\mathrm{m\,s^{-1}}$ \\
Viscosity & $\eta_0$ & 1.1705 & mPa\,s \\
Compressibility & $\kappa_0$ & 440.5 & $\mathrm{TPa^{-1}}$  \\
\end{tabular}
\end{ruledtabular}
\end{table}

\subsection{GDPT setup and method}

The particle trajectories were recorded using the general defocusing particle tracking (GDPT) technique~\cite{Barnkob2015, Barnkob2020, GDPTlab2018}. GDPT is a single-camera particle tracking method, in which astigmatic images are employed by using a cylindrical lens. Such a system provides elliptical shapes of the defocused spherical particles elongated in two perpendicular directions in the depth coordinate $z$, and it constitutes a robust technique for 3D particle tracking in microfluidics. The setup for conducting GDPT measurements consists of an epi-fluorescence microscope (BX51WI, Olympus Corporation, Tokyo, Japan) equipped with a CMOS camera (ORCA-Flash4.0 V3, Hamamatsu Photonics K.K., Japan). An objective lens with 10x magnification and 0.3 numerical aperture was used and a cylindrical lens with a focal length of 300 mm was placed between the camera and the objective at a distance of 20 mm from the camera, which provides a measurement volume of 1.31 $\times$ 1.52 $\times$ 0.15~mm$^3$. The excitation light for the fluorescent particles was provided by a dual-wavelength LED unit (pE-200, CoolLED Ltd., UK) with a peak wavelength of 488 nm. A standard fluorescence filter cube was used with an excitation pass-band from 460~nm to 490~nm and a high-pass emission filter at 520 nm.

To obtain the defocused particle shapes at different height positions, a stack of calibration images for the particles sticking on the channel bottom was recorded with an interval of 1 $\SImum$ in the $z$ coordinate by moving a motorized \emph{z}-stage (MFD, M\"{a}rzh\"{a}user, Wetzlar GmbH \& Co. KG, Wetzlar, Germany) equipped on the microscope. Then, the $z$ position of the stage was fixed and the motion of the particles was recorded. The acquired images were analyzed in GDPTlab that performs normalized cross-correlation to compare the acquired images to the calibration stack. An exposure time of 90~ms and a frame rate of 10~fps were chosen for the measurements of the particles with $d = 0.5$, 1.0, 1.9, and 3.1 $\SImum$, whereas an exposure time of 40~ms and a frame rate of 20~fps were chosen for the measurements of the particles with $d = 4.8~\SImum$. The refractive indices of the liquids required for calculating the true particle position in $z$-coordinate were measured using an automatic refractometer (Abbemat MW, Anton Paar GmbH, Graz, Austria). For the experiments on fluids that had been made spatially inhomogeneous in density and compressibility, the mean value of the refractive indices of the two liquids injected to the channel was used for particle tracking, which gives a maximum error of 1~$\SImum$ in the $z$ direction. Finally, the particle trajectories and velocities were constructed. Particles that had a cross-correlation peak amplitude less than 0.95 and trajectories composed by less than six particle positions were rejected.

\subsection{Experimental procedures}

\begin{figure*}[!t]
\centering
\includegraphics[width=1.0\textwidth]{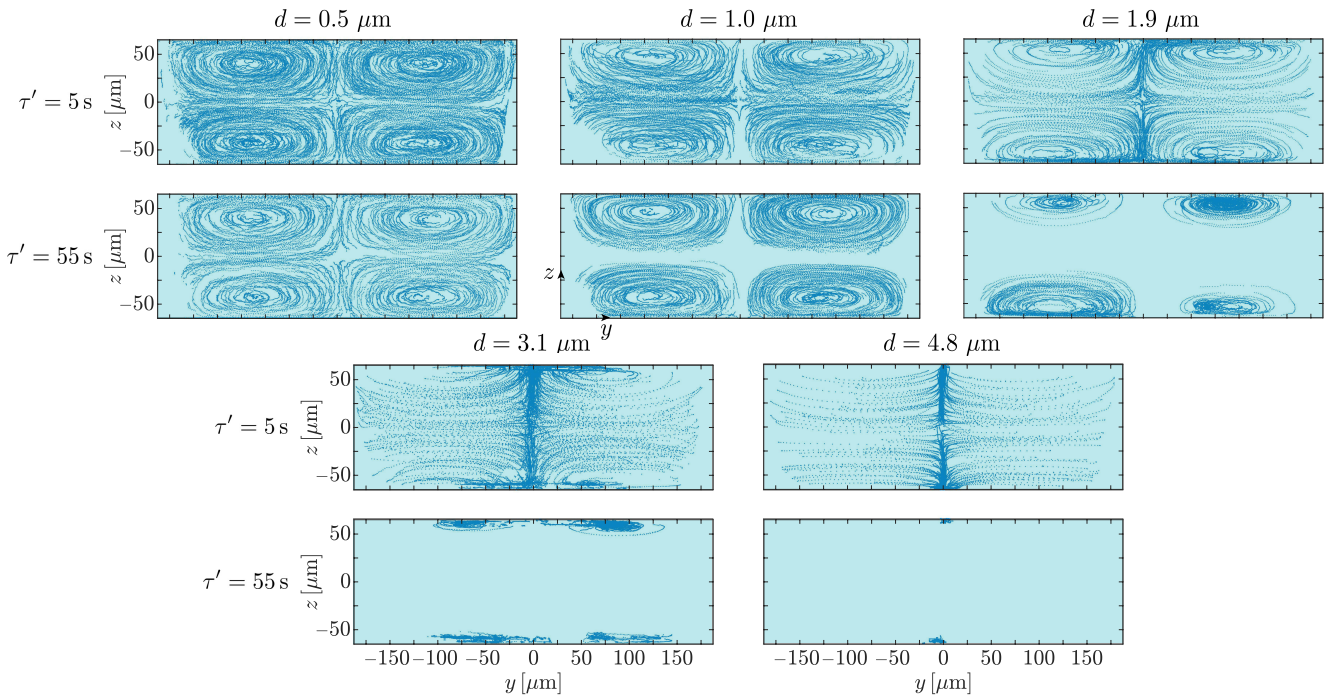}
\caption[]{\label{ParticleTrajectoryHomo}
Experimental particle trajectories in homogeneous medium observed in the vertical $y$-$z$ cross section of width $W=375~\SImum$  and height $H=133~\SImum$ at mid-interval time $\tau^\prime$ = 5 s and 55 s for the particles with different sizes using the 2.5\% Ficoll PM70. Initially, the particles were evenly distributed. Dark blue points are overlaid experimental particle positions from 100 frames for the particles with $d = $ 0.5, 1.0, 1.9, and 3.1 $\SImum$ and from 200 frames for the particles with $d = $ 4.8 $\SImum$ at $\tau^\prime$ = 5 s and 55 s, respectively. The acoustic energy density was set to $E_\mathrm{ac} = 31.4$~Pa.}
\end{figure*}
The liquids were injected in the channel at a total flow rate of 100 $\SImuL$/min, and the volumetric ratio between Milli-Q water and 5\% Ficoll PM70 was close to one in the inhomogeneous case. The PZT transducer was actuated using a linear frequency sweep from 1.95 to 2.05~MHz in cycles of 1~ms to produce a standing half-wave across the width~\cite{Manneberg2009a}. This sweeping range covers the identified resonance frequencies at 1.96 MHz for pure water and 1.97 MHz for 5\% Ficoll PM70 and ensures steady actuation throughout the experiment during the time-evolution of the concentration field. The applied voltage was adjusted to 1.63 V peak-to-peak to achieve the acoustic energy density $E_\mathrm{ac} \approx 31.4$~Pa in the channel. We measured $E_\mathrm{ac}$ in the homogeneous case by tracking individual polystyrene beads with a nominal diameter of 6.33~$\SImum$ (PFP-6052, Kisker Biotech GmbH~\&~Co. KG, Steinfurt, Germany)~\cite{Barnkob2010}, whereas for the inhomogeneous case with three liquid layers, we estimated $E_\mathrm{ac} = \frac12(E_\mathrm{ac}^\mathrm{cntr} + E_\mathrm{ac}^\mathrm{side})$, where $E_\mathrm{ac}^\mathrm{cntr}$ and $E_\mathrm{ac}^\mathrm{side}$ in the center and side layers were measured in their respective homogeneous states. In the homogeneous case, the sound field was turned on simultaneously with the image acquisition, whereas in the inhomogeneous case, the sound field was turned on before stopping the flow in order to stablize the Ficoll solution against gravity~\cite{Augustsson2016,Karlsen2016}. The flow was stopped at time $\tau=0$, and the images for the GDPT measurements were then recorded. The instantaneous stop of the flow in the channel was performed by short circuit of the two inlets through switching a four-port valve, which stops the flows and equilibrates the pressures of the two inlet streams~\cite{qiu2019experimental}. For the measurements using 3.1- and 4.8-$\SImum$-diameter particles in inhomogeneous fluids, the applied voltage was initially set to 0.50 V peak-to-peak for stabilizing the Ficoll solution and then switched to 1.63 V peak-to-peak when the measurements started, which avoids the fast focusing of particles before image acquisition. For each set of measurements, the particle motion was recorded for $60~\mathrm{s}$ in the homogeneous case and $200~\mathrm{s}$ in the inhomogeneous case, to observe the long time-scale evolution of the particle trajectories. Each measurement was repeated at least 16 times to improve the statistics.

\section{Experimental results for homogeneous fluids}
\seclab{exp_homog}

We extend the experimental study of particle-size-dependent particle trajectories to late times of particles suspended in a homogeneous fluid inside a channel with a rectangular cross section embedded in a silicon-glass microchip and exposed to a standing half-wavelength ultrasound wave. We show in Fig.~\ref{ParticleTrajectoryHomo} the trajectories of the particles with diameter $d = 0.5$, 1.0, 1.9, 3.1, and 4.8 $\SImum$ for early (mid-interval time $\tau^\prime = 5$ s) and late ($\tau^\prime = 55$ s) times. Here, mid-interval time $\tau^\prime$ refers to the time interval from $\tau^\prime - 5$ s to $\tau^\prime + 5$ s, and the particle trajectories at each $\tau^\prime$ are plotted by overlaying the particle positions within each time interval. The smallest particles ($d = 0.5~\SImum$) closely follow the boundary-driven Rayleigh streaming from the early time, forming quadrupolar streaming rolls, indicating that the contribution from the acoustic radiation force $\bm{F}^\mathrm{rad}$ is very weak. The particle trajectories at $\tau^\prime = 5$ s closely match those at $\tau^\prime = 55$ s, indicating that the trajectories of 0.5-$\SImum$-diameter particles show no observable time dependency on that time scale.

When the particle diameter increases to 1 $\SImum$, the acoustic streaming is still dominant, and hence the majority of the particles also form quadrupolar rolls at the early time although the vortex size is slightly reduced. The vortex center is also closer to the ceiling and the bottom compared with the trajectories of 0.5-$\SImum$-diameter particles. At late times, the particles deplete along the horizontal and vertical centers and also at the sides of the channel.

\begin{figure*}[!t]
\centering
\includegraphics[width=1.0\textwidth]{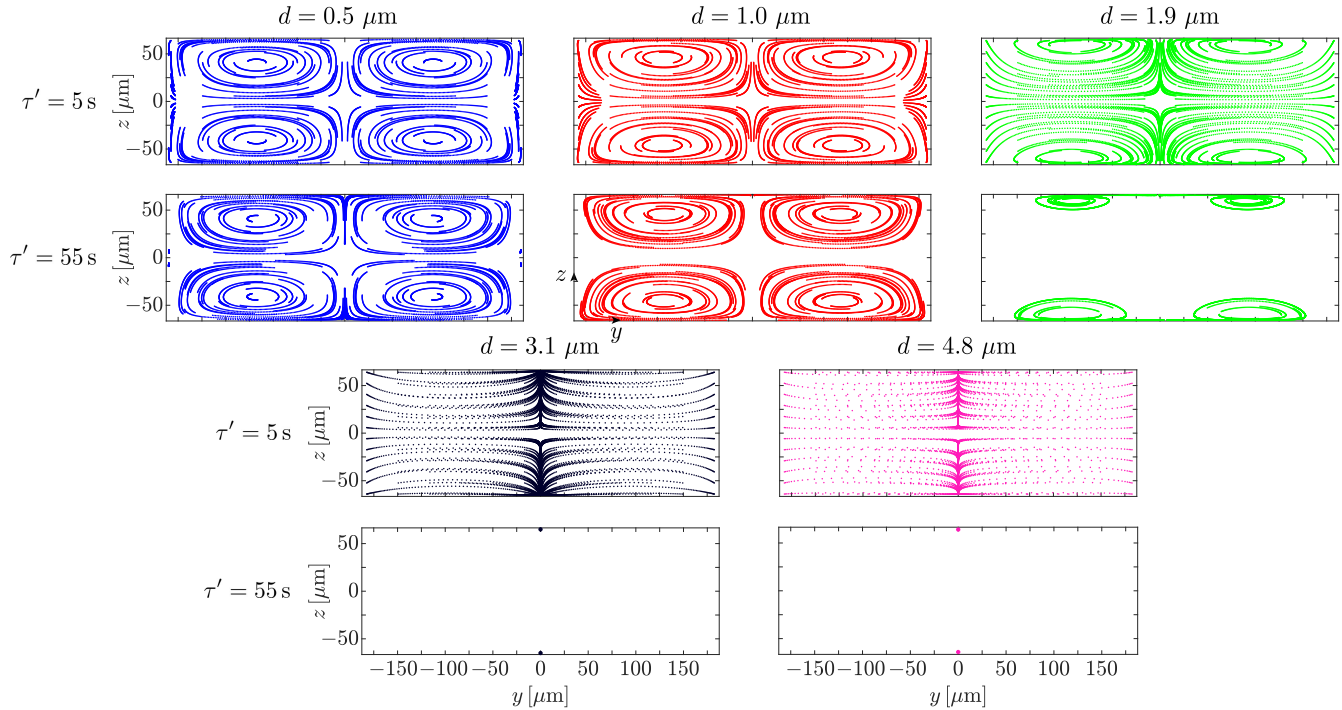}
\caption[]{\label{Particle_trajectory_multi}
Simulated particle trajectories in a homogeneous fluid with $E_\mathrm{ac} = 31.4$~Pa at mid-interval time $\tau^\prime$ = 5 s and 55 s for five different particle diameters, i.e. $0.5\,\SImum$, $1.0\,\SImum$, $1.9\,\SImum$, $3.1\,\SImum$, and $4.8\,\SImum$, which mimic the experimental observations shown in Fig.~\ref{ParticleTrajectoryHomo}. The simulation for each size of particles ran for a physical time $\tau = 60$~s with a time step $\Delta \tau =0.1$ s, and particle positions within 100 time steps were overlaid to form the trajectories at each $\tau^\prime$.}
\end{figure*}

When the particle diameter further increases to 1.9 $\SImum$, a diameter very close to the critical diameter, the particles close to the horizontal center become focused at early times due to the contributions of both $\bm{F}^\mathrm{rad}$ and streaming. The particles close to the ceiling or the floor circulate in four further compacted rolls. All the particles fall into the four rolls at late times, resulting in a relatively large depletion area around the horizontal and vertical centers and at the channel sides.

The particles with a diameter of 3.1 $\SImum$ are considered well above the cross-over size. Thus the $\bm{F}^\mathrm{rad}$ should dominate over streaming and the particles are expected to be focused onto the pressure nodal plane. The measured particle trajectories show that at early times, the vast majority of the particles are indeed focused to the nodal plane and then dragged to the ceiling or the floor due to the streaming. Surprisingly, most of the particles are then further dragged out from the pressure nodal plane and then follow nearly reciprocating paths in very confined regions close to the top and bottom walls at late times. For larger particles ($d = 4.8~\SImum$) which are far above the cross-over size, all of the particles are focused at the initial stage and then stay on the pressure nodal plane, following the streaming, until they are finally gathered in two points at the floor and ceiling.

\section{Numerical results and discussion for homogeneous fluids}
\seclab{discus_homog}

In this section, we study the physical mechanisms of the particle depletion for particles smaller than the cross-over size in homogeneous fluids. We explain it through numerical simulations, and by including the effects of the Brownian motion on particle trajectories.

\begin{figure}[!t]
\centering
\includegraphics[width=1.0\columnwidth]{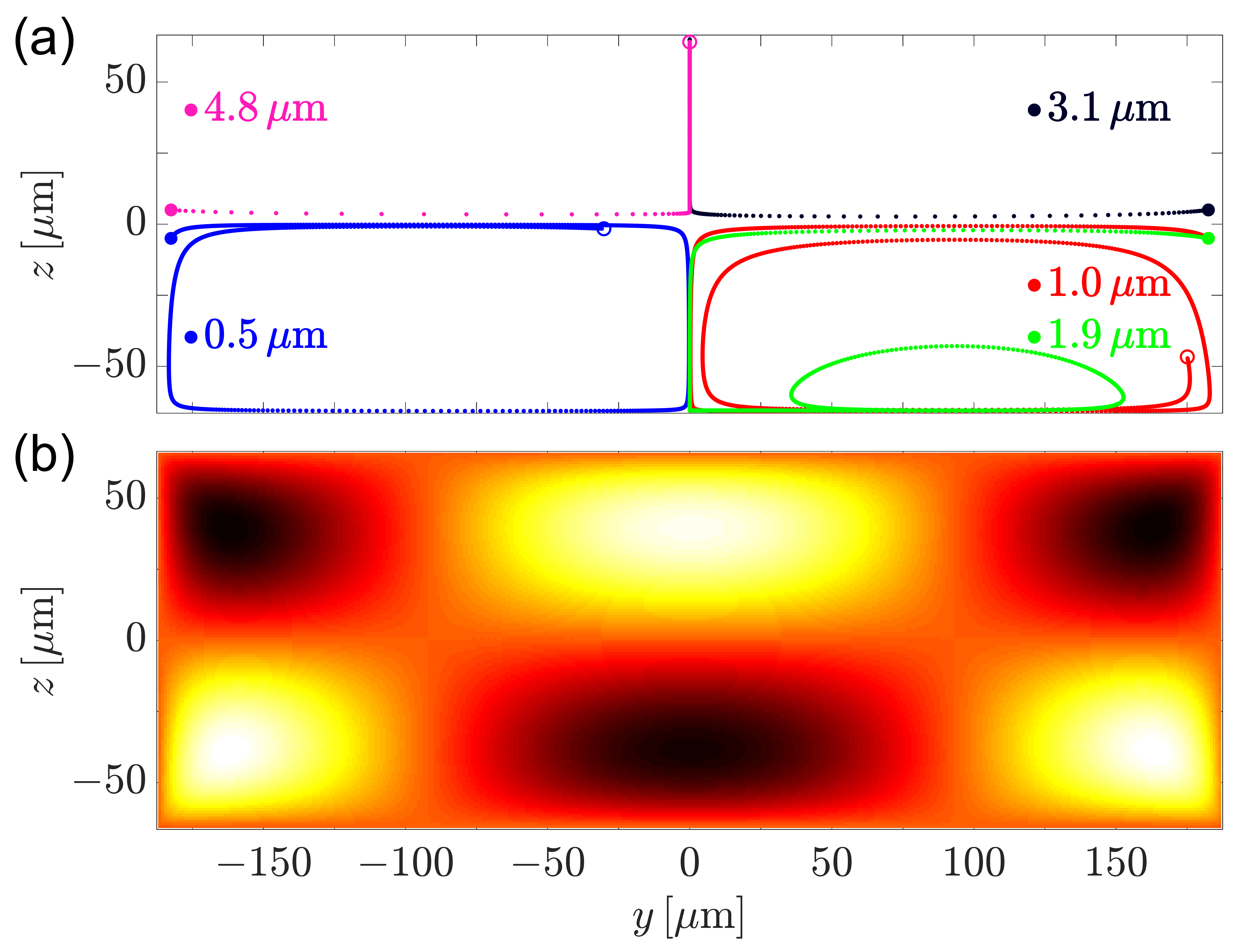}
\caption[]{\label{SingleParticleTrajectory}
Numerical simulations for a homogeneous fluid with $E_\mathrm{ac} = 31.4$~Pa.
(a) Single-particle trajectories within a physical time $\tau = 120$ s. The trajectories of 0.5-, 1.0-, 1.9-, 3.1-, and 4.8-$\SImum$-diameter particles locate in the third, fourth, fourth, first, and second quadrant, respectively. The initial position of each trajectory is indicated by a large solid circle close to the horizontal mid-height plane and to the sidewall, and the end position is indicated by an empty circle. No end position is indicated when the particle trajectory forms a closed loop. (b) The $z$-component of the advective velocity $\big\langle v_{2z} \big\rangle$ in $y$-$z$ cross section. Bright color indicates upward direction whereas dark color indicates downward direction.}
\end{figure}

\subsection{Mechanisms of particle depletion}

To understand the observed particle trajectories, the experimental system was modelled numerically. The particles were modelled as monodisperse spherical polystyrene particles and the particle-particle interaction was neglected. The radiation force $\bm{F}^\mathrm{rad}$ acting on the particles and the acoustic streaming field $\big\langle\bm{v}_2(y,z) \big\rangle$ were calculated numerically using COMSOL Multiphysics in the $y$-$z$ cross section. For simplicity, the wall effect on $\bm{F}^\mathrm{rad}$ and $\big\langle\bm{v}_2(y,z) \big\rangle$ was not taken into account in the model, see Ref.~\cite{Barnkob2012a}. This assumption should be studied in more detail in future work. The effect of gravity was included in all simulations. After extracting the numerical results of $\bm{F}^\mathrm{rad}$ and $\big\langle\bm{v}_2(y,z) \big\rangle$, the fourth-order Runge-Kutta method was applied to compute the particle trajectories from 0 to 120 s in steps of 0.1 s. The accuracy of the fourth-order Runge-Kutta method was checked by going back in time from an end-position of one trajectory to its initial position. The relative error in particle initial position was found to be in the order of $10^{-8}$, which underlines the high accuracy of this method. The energy density was set to $E_\mathrm{ac} = 31.4$ Pa to reflect the experiments. Specular reflection was used when particles hit the walls.

To mimic the experiments, we simulated the trajectories of 144 polystyrene microparticles, suspend in 2.5\% Ficoll PM70 and evenly distributed at time $\tau = 0$ s for each size, Fig.~\ref{Particle_trajectory_multi}. Each plot shows overlaid positions of 144 individual particles within the corresponding time interval. The simulated trajectories have a clear correspondence to the experimental results shown in Fig.~\ref{ParticleTrajectoryHomo}. The trajectories of the particles below the cross-over size ($d = $ 0.5 $\SImum$, 1.0 $\SImum$, and 1.9 $\SImum$) reveal a particle-size-dependent depletion area and particle rolls in good agreement with the experimental results. For motion along $y$, $\bm{F}^\mathrm{rad}$ dominates over streaming everywhere for large particles ($d = 3.1\,\SImum$ and $4.8\,\SImum$). Hence, the particles become focused and stay at the pressure nodal plane and no rolls are formed. The major discrepancy between simulation and experiment is that 3.1-$\SImum$-diameter particles stay on the pressure nodal plane after being focused in simulation, whereas in experiment the majority of the particles are dragged out of the plane by the streaming and form very confined rolls close to the ceiling and the bottom.

To understand the observed particle depletion, we performed a simulation of single particle trajectory for each particle size, Fig.~\ref{SingleParticleTrajectory}(a). The simulation was run for a physical time $\tau = 120$ s, and the initial particle positions were chosen to be close to the horizontal mid-height plane and close to a side wall. The trajectories of 0.5-, 1.0-, and 1.9-$\SImum$-diameter particles form rolls with different sizes. The smallest particle ($d = $ 0.5 $\SImum$) initially follows the outermost stream to the channel bottom, and is then displaced away from the floor near the corner by the streaming $z$-component and forms a slightly smaller roll compared to its initial trajectory. The trajectories of 1.0- and 1.9-$\SImum$-diameter particles follow a similar pattern, but they become displaced at an earlier stage and hence form smaller rolls.

Two factors lead to the early displacement away from the wall for the 1.0- and 1.9-$\SImum$-diameter particles. First, the higher advective $z$-velocity that these larger particles experience because their centers are farther from the wall, and second, the longer retention time of the larger particles at the ceiling and floor. Figure ~\ref{SingleParticleTrajectory}(b) shows that the z-component of the advective velocity $\big\langle v_{2z} \big\rangle$ drops to zero at $z = \pm H/2$. Moreover, particles are dragged towards the ceiling or floor ($z = \pm H/2$) by the streaming field in the region where $-W/4 < y < W/4$, while they are dragged towards the horizontal mid-height plane ($z = 0$) in the region where $\left|W/4 \right| < y < \left|W/2 \right|$. The particles moving along the ceiling or floor depart from the walls when $\big\langle v_{2z} \big\rangle$ points to the horizontal mid-height plane. Since the center of mass of large particles is farther from the walls compared to small particles, they will experience higher $\big\langle v_{2z} \big\rangle$. In addition, large particles move slower along $y$ direction because the velocity components of the radiation force $u^\mathrm{rad}_{y}$ and the streaming $u^\mathrm{str}_{y}$ have opposite directions with close amplitudes. These two effects lead to the earlier displacement away from the wall (\textit{i.e.} closer to $y = \pm W/4$) of large particles than small particles.

\begin{figure*}[!t]
\centering
\includegraphics[width=1.0\textwidth]{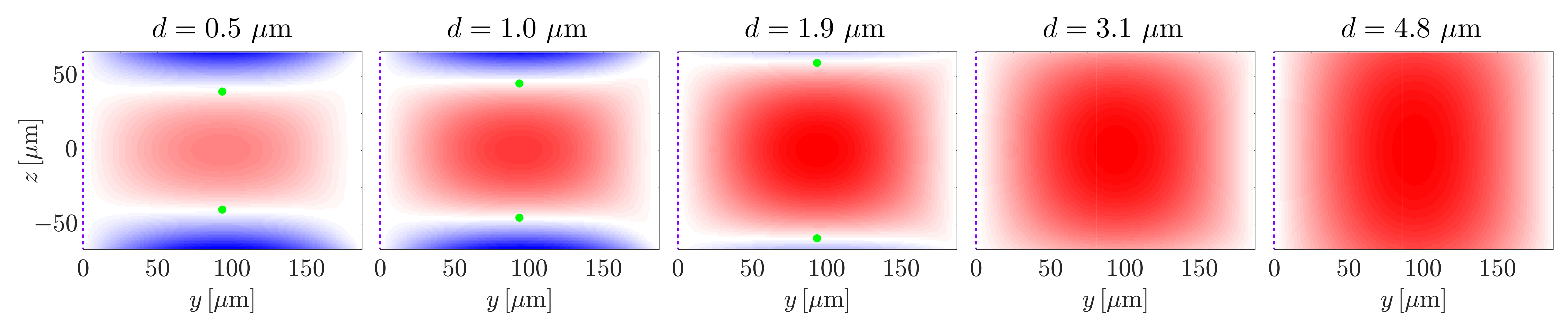}
\caption[]{\label{u^diff_y}
Numerical simulation for a homogeneous fluid with $E_\mathrm{ac} = 31.4$~Pa of the normalized particle velocity difference in the $y$ direction ${u}^\mathrm{diff}_{y}$, see Eq.~(\ref{u^diff_y_equation}), in the right half of the vertical channel cross section from -1 (blue (close to the channel ceiling and bottom), $u^\mathrm{str}_{y}$ dominates) to +1 (red (in the bulk of the channel), $u^\mathrm{rad}_{y}$ dominates), through 0 (white, balance $u^\mathrm{str}_{y} = u^\mathrm{rad}_{y}$). The center positions of the rolls in the trajectories of 0.5, 1.0, and 1.9-$\SImum$-diameter particles at late times are indicated by solid light green points. Only half of the channel cross section is shown since ${u}^\mathrm{diff}_{y}$ is symmetric with respect to the pressure nodal plane.}
\end{figure*}

Once the particles have been displaced away from the floor or ceiling, they will turn towards the pressure nodal plane when $u^\mathrm{rad}_{y}$ dominates over $u^\mathrm{str}_{y}$ in the bulk. When 1.9-$\SImum$-diameter particles start to move towards the pressure nodal plane, they soon experience large $\bm{F}^\mathrm{rad}$ pushing them quickly from the upward $\big\langle v_{2z} \big\rangle$ region ($\left|W/4 \right| < y < \left|W/2 \right|$) to the downward $\big\langle v_{2z} \big\rangle$ region ($-W/4 < y < W/4$), which eventually form more confined rolls and larger depletion areas compared to the cases for 0.5- and 1.0-$\SImum$-diameter particles. The particle depletion is strongly correlated with the competition between $u^\mathrm{rad}_{y}$ and $u^\mathrm{str}_{y}$. The difference between the two velocity components $u^\mathrm{diff}_{y}$ is defined as
\begin{equation}
\label{u^diff_y_equation}
u^\mathrm{diff}_{y} = u^\mathrm{rad}_{y} - u^\mathrm{str}_{y},
\end{equation}
which is plotted in Fig.~\ref{u^diff_y}. It is found that the positions of the roll centers formed by 0.5-, 1.0-, and 1.9-$\SImum$-diameter particles at late times are located at $y = \pm W/4$, $z = 26.5\,\SImum$, $21\,\SImum$, and $7\,\SImum$, respectively, where ${u}^\mathrm{diff}_{y} = 0$. It is also found that the motion of 0.5- and 1.0-$\SImum$-diameter particles at the two sides of the rolls is not symmetric with respect to $y=\pm W/4$, resulting in an inwards spiral motion. This is because the particles close to the sidewalls move more slowly along $z$ direction than the particles moving close to $y=0$, owing to the no-slip boundary condition on the sidewalls (Fig.~\ref{SingleParticleTrajectory}(b)). Hence the $\bm{F}^\mathrm{rad}$ close to the sidewalls has a longer time to push the particles inwards, compared with the time that $\bm{F}^\mathrm{rad}$ has to push the particles outwards when they move close to the pressure nodal plane. The inwards spiral motion occurs because the particles are not being pushed outwards, when they are near the channel horizontal center, as much as they are pushed inwards, when they are close to the sidewalls. This also contributes to the formation of the depletion area.

Perhaps counter-intuitively, gravity also influences the trajectories of in particular the 1.9-$\SImum$-diameter particles. The particles in the upper half of the plane move away from the ceiling faster compared to the particles in the lower half, since gravity points downwards. This effect results in slightly smaller rolls formed by the particles in the upper half and bigger rolls in the lower half (Fig.~\ref{Particle_trajectory_multi}). This is also the reason that the 1.9-$\SImum$-diameter particle in Fig.~\ref{SingleParticleTrajectory} forms a closed loop. The inward motion is compensated by gravity and hence the particle always reaches the channel bottom. It should be noted that the depletion area is also related to the energy density $E_\mathrm{ac}$. Under higher $E_\mathrm{ac}$, the depletion area will grow faster compared with the case under low $E_\mathrm{ac}$, because the particles will travel more loops within a certain time due to their high velocity.

The particles with large sizes ($d = $ 3.1 $\SImum$ and 4.8 $\SImum$) are focused to the pressure nodal plane, after which they are dragged to the ceiling or floor and finally remain there without being dragged out. The streaming does not play a major role in this case, since $u^\mathrm{rad}_{y}$ dominates over $u^\mathrm{str}_{y}$ everywhere in the channel (Fig.~\ref{u^diff_y}).

There is a notable discrepancy between the trajectories of 3.1-$\SImum$-diameter particles in experiment (Fig.~\ref{ParticleTrajectoryHomo}) and simulation (Fig.~\ref{Particle_trajectory_multi}), which may be attributed to the omission in the model of a near-wall Stokes drag correction. The wall corrections on the drag force have been investigated by Fax{\'e}n for spherical particles moving parallel to the wall~\cite{Faxen1922} and by Brenner~\cite{Brenner1961} for spherical particles moving perpendicular to the wall (a thorough review can be found in Ref.~\cite{Happel1983}). The parallel wall correction factor for the particles moving on the wall is around 1.9, but drops to 1.3  just three particle radii away from the wall~\cite{Koklu2010}. The transition from the regime where particles form rolls to the regime where particles stay in the horizontal center ($y = 0$) at late times is sharp, meaning that the particles with a diameter difference of 0.1 $\SImum$ can belong to two different regimes, owing to the tiny ${u}^\mathrm{diff}_{y}$ when the particle size is very close to the cross-over size. The cross-over particle diameter $d_\mathrm{c}$ in simulation will be larger than the current one ($d_\mathrm{c} = 2.3\,\SImum$), if the wall corrections are included in the model, which will be much closer to the $d_\mathrm{c}$ in experiment.

\begin{figure*}[!t]
\centering
\includegraphics[width=1.0\textwidth]{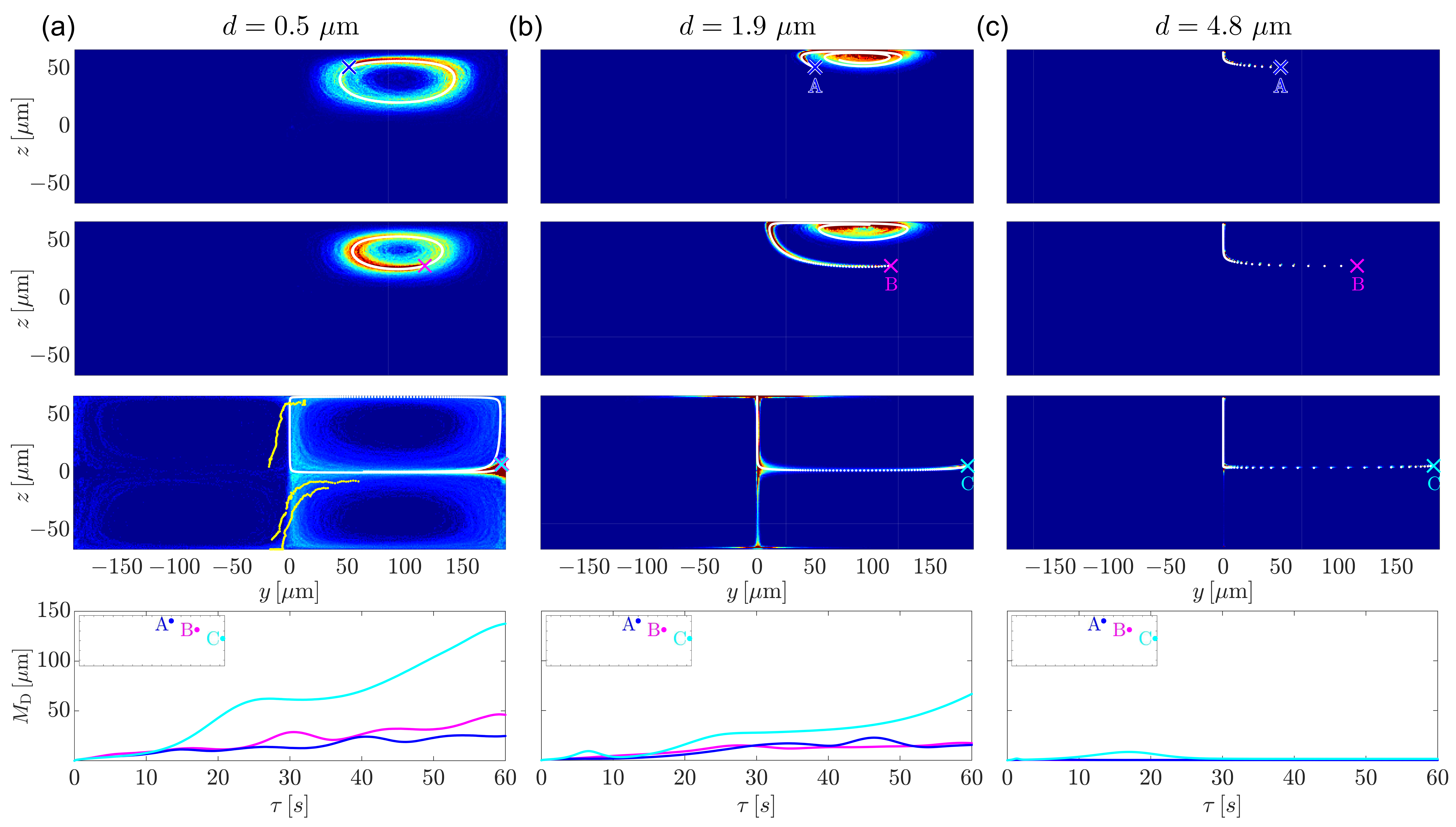}
\caption[]{\label{Brownian_motion}
Simulations of particle trajectories for a homogeneous fluid with $E_\mathrm{ac} = 31.4$~Pa taking into account the Brownian motion plotted together with the deterministic trajectory (white) within a physical time $\tau = 60$ s for the particles with $d$ = (a) 0.5 $\SImum$, (b) 1.9 $\SImum$, and (c) 4.8 $\SImum$. The initial particle position for each plot is indicated by a cross. The initial positions in the first, second, and third rows are named as A, B, and C, which are indicated in the insets of the plots in the last row. For each initial particle position, 1000 simulations were performed and the color plot indicates the particle population from low (light blue) to high (dark red). For a 0.5-$\SImum$-diameter particle starting from position C, simulations were performed for $\tau = 120$ s in order to show 0.5-$\SImum$-diameter particle starting from the first quadrant can form rolls in any of the four quadrants. The maximum population varies in each plot, and hence the range of the color plot is not consistent which can only be used for indicating the population in each plot. The experimental particle trajectories showing the switch of the quadrants are indicated by light yellow in the third row of (a). The last row shows the mean value of the particle distance $M_\mathrm{D}$ between each single trajectory and the deterministic trajectory as a function of $\tau$.}
\end{figure*}

\subsection{Effect of Brownian motion on the particle trajectories}

Since the microparticle transport is very sensitive to small perturbations, we investigated numerically the effect of Brownian motion on the microparticle trajectories. The 2D particle displacement in space $\Delta s_\tau$ in our system is described by the stochastic differential equation
\begin{equation}
\Delta s_\tau = v(s,\tau)\Delta \tau + \sqrt{2D}\,Z\, \sqrt{\Delta \tau},
\end{equation}
where $D$ and $Z$ are the diffusion coefficient of microparticles (see Table~\ref{MaterialParameters}) and the standard normal variable.

\begin{figure*}[!t]
\centering
\includegraphics[width=1.0\textwidth]{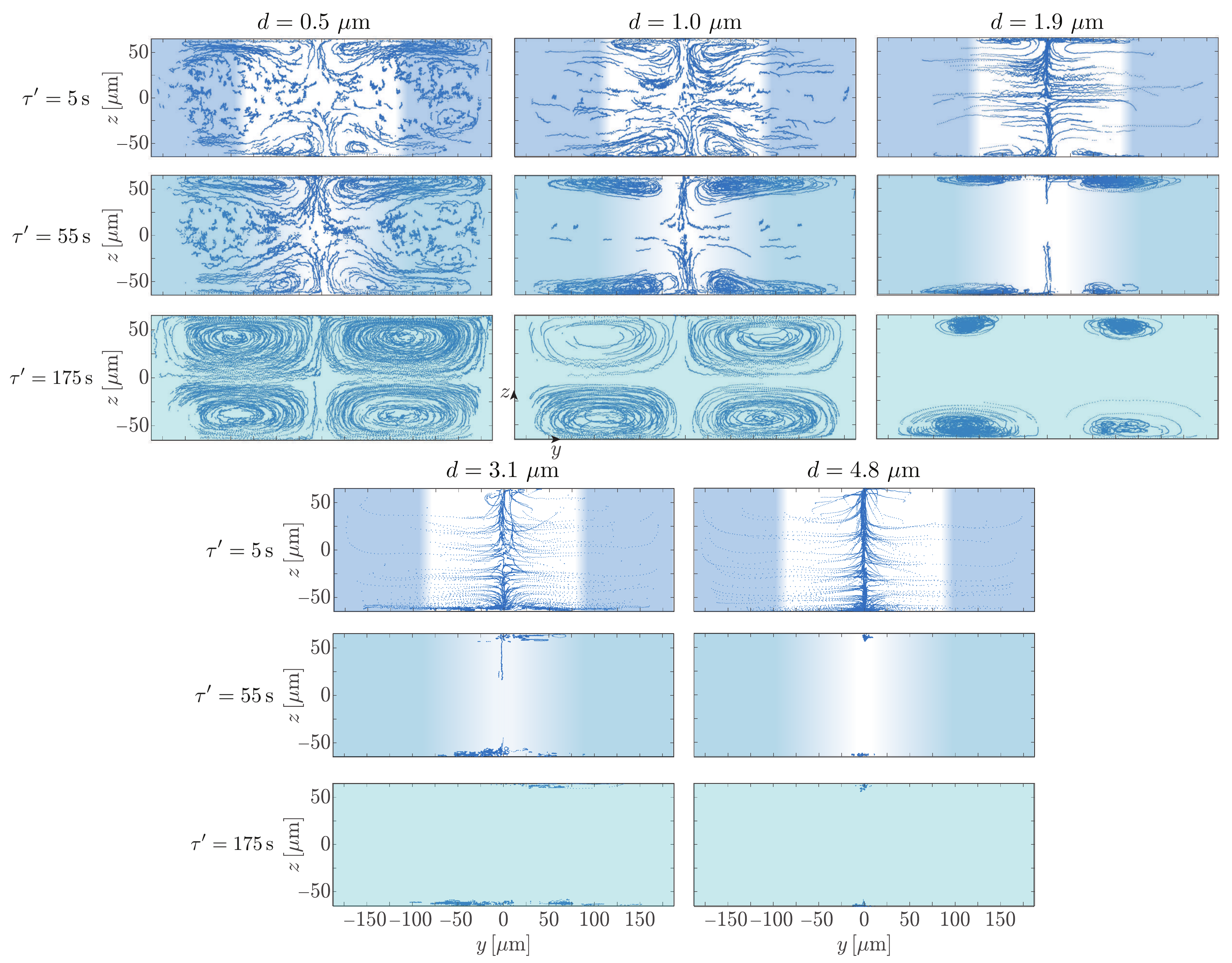}
\caption[]{\label{Particle_trajectory_inhomo}
Particle trajectories in an inhomogeneous fluid with $E_\mathrm{ac} = 31.4$~Pa measured in the vertical $y$-$z$ cross section of the microchannel at mid-interval time $\tau^\prime$ = 5 s, 55 s, and 175 s for particles of different sizes in inhomogeneous solutions (center stream: 5\% Ficoll PM70; side stream: Milli-Q water). Dark blue points are overlaid experimental particle positions from 100 frames for the particles of $d = $ 0.5 $\SImum$, 1.0 $\SImum$, 1.9 $\SImum$, and 3.1 $\SImum$ and from 200 frames for the particles of $d = $ 4.8 $\SImum$ at $\tau^\prime$ = 5 s, 55 s, and 175 s, respectively. The color plot represents the concentration of the solute molecules from low (dark) to high (white).}
\end{figure*}

The trajectories of three sizes of particles ($d$ = 0.5 $\SImum$, 1.9 $\SImum$, and 4.8 $\SImum$) are shown in Fig.~\ref{Brownian_motion} for the three initial particle positions in the first quadrant. For each initial position, 1000 repeated simulations were performed for a physical time $\tau = 60$ s (for a 0.5-$\SImum$-diameter particle starting from position C, simulations were performed for $\tau = 120$ s). For 0.5-$\SImum$-diameter particles that start out within the inner stream, all the particle trajectories fall into the first quadrant with a wide distribution around the deterministic particle trajectory (white). The mean distance $M_\mathrm{D}$ between each individual trajectory and the deterministic trajectory increases with $\tau$. When the particle starts the trajectory close to the boundary of the quadrant, its trajectory may end in any of the four quadrants since the particle displacement due to the Brownian motion is sufficiently large for the particle to switch the quadrant, which has also been observed in experiments (yellow overlaid trajectories). This phenomenon results in a much larger $M_\mathrm{D}$ at late times. When the particle size increases, the range of the particle trajectory distribution reduces and individual trajectories are closer to the deterministic one, since $D$ decreases with increasing particle size (Table~\ref{MaterialParameters}) and hence the particle displacement due to the Brownian motion is small. The trajectories of 4.8-$\SImum$-diameter particles closely follow the deterministic trajectories for all the three initial positions, leading to small $M_\mathrm{D}$ at all times. The simulations indicate that Brownian motion is an important factor for particle transport in acoustophoresis for the particles below the cross-over size.

\section{Experimental results and discussion for inhomogeneous fluids}
\seclab{exp_inhomog}

Finally, we turn to the case of fluids that are made spatially inhomogeneous in density and compressibility. We study experimentally the corresponding time- and particle-size-dependent particle trajectories from early to late times using the same device as above. The resulting particle trajectories are shown in Fig.~\ref{Particle_trajectory_inhomo}. Because acoustic streaming is substantially suppressed in the bulk of the channel at early times~\cite{Karlsen2018, qiu2019experimental}, the trajectories of small particles are radically different from those in homogeneous medium. The 0.5-$\SImum$-diameter particles close to the ceiling and the floor follow flat streaming rolls at early times, while the particles in the bulk experience a weak radiation force $\bm{F}^\mathrm{rad}$ that slowly pushes them towards the pressure nodal plane. As time evolves, the streaming rolls grow due to both diffusion and advection~\cite{qiu2019experimental}, and hence, more particles are dragged into the rolls. After the medium becomes homogeneous at late times, the four quadrupolar streaming rolls appear, which are clearly distinguishable in the case of 0.5-$\SImum$-diameter particles. The 1.0- and the 1.9-$\SImum$-diameter particles experience stronger $\bm{F}^\mathrm{rad}$, so the particles in the bulk focus to the pressure nodal plane already at early times. However, particles that are close to the ceiling or floor still form confined rolls due to the strong streaming near the horizontal boundaries. After the inhomogeneity smears out at late times, the particles form four rolls and a depletion area similar to the case in homogeneous fluids. The trajectories of 1.0-$\SImum$-diameter particles were previously studied in a gradient created by iodixanol and Milli-Q in Ref.~\cite{Karlsen2018} at early and intermediate times, but the particle trajectories at late time were not reported and hence the particle depletion was not discovered. The majority of 3.1-$\SImum$-diameter particles are focused to the horizontal center by $\bm{F}^\mathrm{rad}$ at early times, but are then dragged out by the streaming and move in a flat and very confined region, corresponding to the situation in homogeneous medium. 4.8-$\SImum$-diameter particles are quickly focused to the pressure nodal plane and remain trapped already at early times, due to the dominating $\bm{F}^\mathrm{rad}$. The particle-size-dependent particle trajectories in inhomogeneous fluids show distinct difference for small particles at early times compared with those in a homogeneous fluid. The majority of the particles can be focused by $\bm{F}^\mathrm{rad}$ due to the streaming suppression in inhomogeneous medium, which is not possible in the homogeneous case. This new feature opens up the possibility to focus and separate the particles below the critical size, which has recently been demonstrated~\cite{Assche2020}.

\section{Conclusions}
\seclab{conclusion}

Using the GDPT technique, we have measured the 3D acoustophoretic motion at long time scales of microparticles suspended in either homogeneous or inhomogeneous fluids, and characterized the cross-over from radiation force dominated to streaming dominated motion for decreasing particle size. Here, long time scales means that the particles smaller than the cross-over size move over distances comparable to the channel width.

For homogeneous fluids, we observed both experimentally and in numerical simulations that the particles below the cross-over size form depletion regions at late times, and that those regions increase as the particle size increases. We found good agreement between experimental measurements and numerical simulations of the particle motion. Through the numerical simulations, we found that the depletion is partially due to the displacement at early times away from the floor and ceiling of the channel for large particles. This confines the large particles to smaller flow rolls than the small particles. The particle depletion is also due to the inwards spiral motion caused by the slow particle motion in the $z$ direction near the sidewalls resulting from the no-slip boundary condition. The effect of Brownian motion was also investigated in the numerical simulations, and it was found to be important for the motion of particles with sizes below the critical size.

For inhomogeneous fluids, the experimentally observed particle trajectories exhibit focusing in the bulk of the microchannel at early times even for particles below the cross-over size due to a remarkable suppression of acoustic streaming, which shows a clear potential for manipulation of submicrometer particles. We did not perform numerical simulations for inhomogeneous fluids.

This work has characterized the motion of suspended particles in acoustofluidic devices as a function of particle size. The  results can be used as a guideline for particle-size-sensitive acoustophoretic particle manipulations in both homogeneous and inhomogeneous fluids.

\section{Acknowledgements}

The idea of this study arose from the initial observation of the depletion of 1.0-$\SImum$-diameter particles in flow experiment performed by D. Van Assche (Lund University). We are grateful to J. S. Bach (Technical University of Denmark) and F. Garofalo (Lund University) for helpful comments and suggestions regarding numerical simulations, R. Barnkob (Technical University of Munich) and M. Rossi (Technical University of Denmark) for providing the software GDPTlab, and Jeppe Revall Frisvad (Technical University of Denmark) for accessing the refractometer. \\[-3mm]

W.Q. was supported by the People Programme (Marie Curie Actions) of the European Union's Seventh Framework Programme (FP7/2007-2013) under REA grant agreement no. 609405 (COFUNDPostdocDTU), and by the Foreign Postdoctoral Fellowship from Wenner-Gren Foundations. H.B. was supported by the Villum Foundation Grant no. 00022951. P.A. was supported by Grant no. 2016-04836 from the Swedish Research Council, Grant no. ICA16-0002 from the Swedish Foundation for Strategic Research, and Grant no. 20180837 from the Crafoord Foundation.

%
%


%

\end{document}